\documentstyle[12pt]{article}

\begin{document}
\title{Breakdown of the Mechanism of Forming Wakes by a Current-Carrying String}
\author{A. L. N. Oliveira$^{1}$ and M. E. X. Guimar\~aes$^{2}$  \\
\mbox{\small{1. Instituto de F\'{\i}sica, Universidade de Bras\'{\i}lia }} \\
\mbox{\small{2. Departamento de Matem\'atica, Universidade de Bras\'{\i}lia}} \\
\mbox{\small{\bf  andreo@fis.unb.br, emilia@mat.unb.br}}}
\maketitle
\begin{abstract}
In this letter we emphasize the effect that the inclusion of 
electromagnetic properties 
for a string 
brings logarithmic divergences to the accretion problem and the 
mechanism of formation and 
evolution of wakes can break down. 
\end{abstract}

\section{Introduction}
Topological defects are predicted in many gauge models as solitonic 
solutions resulting from spontaneous breaking of gauge or global symmetries. 
Among all these solutions, cosmic strings have attracted attention because 
they may be the source of large-scale structure in the Universe~\cite{vil}. 
A relevant mechanism to understand the structure formation by cosmic strings 
involves long strings moving with relativistic speed in the normal plane, 
giving rise to velocity perturbations in their wake~\cite{silk}. 
The mechanism of forming wakes has been considered by many authors in the 
General Relativity theory~\cite{vacha1,vacha2}. More recently, Masalskiene and 
Guimar\~aes~\cite{sandra} have considered the  formation of wakes by long 
strings in the contexte of scalar-tensor theories. They have shown that the 
presence of a gravitational scalar field - which from now on we will call 
generically as ``dilaton" -  induces a very similar structure as in the case 
of a wiggly cosmic string~\cite{vacha2}. Further, Bezerra and 
Ferreira~\cite{bezerra} showed that, if a torsion is presented, this 
effect is amplified.

From the point of view of purely gravitational physics, it was shown that the 
inclusion of a current hardly affects the metric outside the 
string~\cite{peter2}. 
However, it is well-known that inclusion of such an internal structure 
can change the 
predictions of cosmic string models in the microwave background 
anisotropies~\cite{peter3,peter4}. This is the main goal of this work. 
Namely, we 
analyse the  formation and evolution of wakes in this spacetime and we show 
explicitly how the current affects this mechanism. For this purpose, we 
consider a 
model in which non-baryonic cold dark matter propagates around the 
current-carrying 
string. The Zel'dovich approximation is carried out in order to treat 
this motion. We 
anticipate that our main result is to show that the inclusion of a 
current brings 
logarithmic divergences and can actually break down the accretion 
mechanism by wakes. We take here a 
slightly different approach from papers \cite{others}, although our 
results are very similar 
of them.  

This work is outlined as follows. In section 2, after setting the relevant 
microscopic model which describes a superconducting string carrying a 
current of timelike type, we present its  gravitational field. 
In section 3, we consider the mechanism of formation and evolution of 
wakes in this 
framework by means of the Zel'dovich approximation. We obtained our results 
in the weak-field approximation.
Finally, in section 4, we end up with some conclusions and remarks.

\section{Timelike Current-Carrying Strings:}

In this section we will study the gravitational field generated by a 
string carrying a current
of timelike-type. We start with the action in the Jordan-Fierz frame
\begin{equation}
\label{acao_JF}
 {\cal{S}} = \frac{1}{16 \pi} \int d^4 x \sqrt{-\tilde{g}} 
[ \tilde{\Phi} \tilde{R} - \frac{\omega(\tilde{\Phi})}{\tilde{\Phi}}
\tilde{g} ^{\mu \nu} \partial _{\mu} \tilde{\Phi} \partial _{\nu} 
\tilde{\Phi} ] + {\cal{S}} _{m} [\psi _m, \tilde{g} _{\mu \nu}]
\end{equation}
$\tilde{g}_{\mu\nu}$ is the physical metric which contains both scalar
and tensor degrees of freedom, $\tilde{R}$ is the curvature scalar
associated to it and ${\cal S}_{m}$ is the action for general matter
fields which, at this point, is left arbitrary. The metric signature
is assumed to be $(-,+,+,+)$.

In what follows, we will concentrate our attention to superconducting 
vortex configurations which arise from the spontaneous breaking of the 
symmetry $U(1) \times U_{em}(1)$. Therefore,
the action for the matter fields will be composed by two pairs of coupled 
complex scalar and gauge fields $(\varphi, B_{\mu})$ and $(\sigma, A_{\mu})$.
Also, for technical purposes, it is preferable to work in the so-called 
Einstein (or conformal) frame, in which the scalar and tensor degrees of 
freedom do not mix.
\begin{eqnarray}
{\cal{S}}&=& \frac{1}{16 \pi G_*} \int d^4x \sqrt{-g} [R-2\partial _{\mu} 
\phi  \partial^{\mu} \phi ] \nonumber \\
&+& \int d^4x \sqrt{-g} [ -\frac{1}{2}\Omega ^2 (\phi) 
((D_{\mu}\varphi) ^{*} D^{\mu} \varphi + 
(D_{\mu} \sigma) ^{*} D^{\mu} \sigma) \nonumber \\
&-& \frac{1}{16 \pi}  ( F_{\mu \nu} F^{\mu \nu} + 
H_{\mu \nu} H^{\mu \nu}) - \Omega ^2(\phi)  V(|\varphi|,|\sigma|) ] ,
\end{eqnarray}
where
$F_{\mu \nu} = \partial{\mu} A_{\nu} - \partial _{\nu} A_{\mu}$ ,
$H_{\mu \nu }= \partial{\mu} B_{\nu} - \partial _{\nu} B_{\mu}$
and the potential is suitably chosen in order that the pair 
$(\varphi, B_{\mu})$ breaks one symmetry $U(1)$ in vacuum 
(giving rise to the vortex configuration) and the second pair 
$(\sigma, A_{\mu})$ breaks the symmetry $U_{em}(1)$ in the core of 
the vortex (giving rise to the superconducting properties)
\begin{equation}
V(|\varphi|,|\sigma|)= \frac{\lambda _{\varphi}}{8} 
(|\varphi|^2- \eta ^2)^2 + f(|\varphi|^2 - \eta ^2) |\sigma|^2 + 
\frac{\lambda _{\sigma}}{4}
|\sigma|^4 + \frac{m_{\sigma}^2}{2} |\sigma|^2
\end{equation}

We restrict, then, ourselves to the configurations corresponding to an 
isolated, static current-carrying vortex lying in the $z-axis$. In a 
cylindrical coordinate system $(t,r,z,\theta)$ such
that $r \geq 0$ and $0 \leq \theta < 2\pi$, we make the following ansatz:
\begin{equation}
\varphi = \varphi(r)e^{i\theta} \;\;\;\;\; B_{\mu}= \frac{1}{q}[Q(r)-1]
\end{equation}
The pair $(\sigma, A_{\mu})$, which  is responsible for the 
superconducting properties of the vortex,  is set in the form
\begin{equation}
\sigma= \sigma(r)e^{i\psi(t)} \;\;\;\;\; A_t = 
\frac{1}{e}[P_t(r)-\partial_t \psi]
\end{equation}
where $P_t$ corresponds to the electric field which leads to a 
timelike current in the vortex. We also require that the 
functions $\varphi, Q(r), \sigma(r) \mbox{and} P_t$ must be 
regular everywhere and must satisfy the usual boundary conditions 
of vortex \cite{ole} and superconducting configurations \cite{witten,carter}.

The action (2) is obtained from (\ref{acao_JF}) by a conformal transformation
\begin{eqnarray*}
\tilde{g} _{\mu \nu} = \Omega ^2(\phi) g_{\mu \nu}
\end{eqnarray*}
and by the redefinition of the quantity
\begin{eqnarray*}
G_* \Omega ^2(\phi)= \tilde{\Phi} ^{-1}
\end{eqnarray*}
which makes evident the feature that any gravitational phenomena 
will be affected by the variation
of the gravitation constant $G_*$ in the scalar-tensor gravity, 
and by introducing a new parameter
\begin{eqnarray*}
\alpha^2 = ( \frac{\partial \ln\Omega(\phi)}{\partial \phi} )^2=
 [ 2 \omega (\tilde{\Phi}) +3 ] ^{-1}
\end{eqnarray*}

Variation of the  action (2) with respect to the metric $g_{\mu\nu}$ 
and to the dilaton field $\phi$ gives the modified Einstein's equations  
and a wave equation for the dilaton, respectively. Namely,
\begin{eqnarray}
\label{EE0}
G_{\mu \nu} &=& 2 \partial _{\mu} \phi \partial _{\nu} \phi -
g_{\mu \nu} g^{\alpha \beta} \partial _{\alpha}
\phi \partial _{\beta} \phi +8 \pi G_*T_{\mu \nu} \nonumber \\
\Box _{g} \phi &=& -4\pi G_*\alpha(\phi)T
\end{eqnarray}

Where $T_{\mu\nu}$ is the energy-momentum tensor which is obtained by
\begin{equation}
\label{E-M}
T_{\mu \nu} = \frac{-2}{\sqrt{-g}} \frac{\delta S_{mat}}{\delta g_{\mu\nu}} .
\end{equation}
We note, in passing, that, in the conformal frame, this tensor is not 
conserved providing us with an additional equation: $
\nabla_{\mu} T ^{\mu}_{\nu}=\alpha (\phi)T\nabla _{\nu} \phi .$

In what follows, we will write the general static metric with 
cylindrical symmetry corresponding to the electric case in the form
\begin{equation}
\label{metrica1}
ds^2 = - e^{2\Psi}dt^2 +
 e^{2(\gamma - \Psi)}(dr^2 + dz^2) + \beta^2 e^{-2\Psi}d\theta^2
\end{equation}
where $\Psi, \gamma, \beta$ are functions of $r$ only. 

The non-vanishing components of the energy-momentum tensor 
using (\ref{E-M}) are
\begin{eqnarray}
T^t_t &=& -\frac{1}{2} \Omega ^2(\phi) \left\{ -g^{tt} \sigma ^2 
{P_t}^2 + g^{rr} \left[{\varphi ^\prime}^2 +{\sigma ^\prime}^2\right]+ 
g^{\theta \theta}\varphi ^2Q^2 \right\} \nonumber \\ 
&-& \frac{1}{8\pi}  g^{rr} \left\{-g^{tt}
\frac{{P_t^\prime}^2}{e^2} + g^{\theta \theta} 
\frac{{Q^\prime}^2}{q^2} \right\} - 
\Omega ^4(\phi) V(\sigma,\varphi) \nonumber \\
T^r_r&=& -\frac{1}{2} \Omega ^2(\phi) \left\{ g^{tt} \sigma ^2 
{P_t}^2 - g^{rr} \left[{\varphi ^\prime}^2 +{\sigma ^\prime}^2\right]+ 
g^{\theta \theta}
\varphi ^2Q^2\right\} \nonumber \\
&+& \frac{1}{8\pi}  g^{rr} \left\{ g^{tt}
\frac{{P_t^\prime}^2}{e^2} + g^{\theta \theta} \frac{{Q^\prime}^2}{q^2} 
\right\} - \Omega ^4(\phi) V(\sigma,\varphi) \nonumber \\
T^{\theta}_{\theta}&=& -\frac{1}{2} \Omega ^2(\phi) 
\left\{ g^{tt} \sigma ^2 {P_t}^2 + g^{rr} \left[{\varphi ^\prime}^2 
+{\sigma ^\prime}^2\right]- g^{\theta \theta}
\varphi ^2Q^2 \right\} \nonumber \\
&-& \frac{1}{8\pi}  g^{rr} \left\{ g^{tt}
\frac{{P_t^\prime}^2}{e^2} - g^{\theta \theta} 
\frac{{Q^\prime}^2}{q^2} \right\} - \Omega ^4(\phi) V(\sigma,\varphi) \\
T^z_z&=& -\frac{1}{2} \Omega ^2(\phi) \left\{ g^{tt} 
\sigma ^2 {P_t}^2 + g^{rr} \left[{\varphi^\prime}^2 +
{\sigma^\prime}^2\right]+ g^{\theta \theta}
\varphi ^2Q^2 \right\} \nonumber \\ 
&-& \frac{1}{8\pi}  g^{rr} \left\{ g^{tt}
\frac{{P_t^\prime}^2}{e^2} + g^{\theta \theta} \frac{{Q^\prime}^2}{q^2} 
\right\} - \Omega ^4(\phi) V(\sigma,\varphi). \nonumber 
\end{eqnarray}

Therefore, for the electric case, eqs. (6) are written as 
\begin{eqnarray}
\label{EE1}
\beta ^{\prime \prime} &=& 8 \pi G_{*} e^{2(\gamma -\psi)} 
\beta [T_1^1+T_3^3] \nonumber \\
(\beta \psi ^{\prime})^{\prime} &=& 4 \pi G_{*} 
e^{2(\gamma - \psi)} \beta [-T_0^0+T_1^1+T_2^2+T_3^3] \nonumber \\
\beta ^{\prime} \gamma ^{\prime}&=& \beta (\psi ^{\prime})^2 + 
\beta (\phi ^{\prime})^2 + 8 \pi G_{*} \beta e^{2(\gamma - \psi)}T_1^1 
\nonumber \\
(\beta \phi ^{\prime})^{\prime} &=& -4 \pi G_{*} \beta e^{2(\gamma -\psi)}T
\end{eqnarray}

In order to solve the above equations we will divide the space in two regions: 
the exterior region, $r \geq r_0$, in which only the electric component of the 
Maxwell tensor contributes to the energy-momentum tensor and the 
internal region, 
$0 \leq r < r_0$, where all matter fields survive. $r_0$ is the 
string thickness. In 
the paper~\cite{nos}, the metric of such a configuration has been obtained to 
first order in the parameter $G_0$. In order to obtain this metric, we used a 
method applied first by Linet \cite{lin} which consists in re-writing 
the components 
of the energy-momentum tensor (9) as $\delta$-functions. We have then
\begin{eqnarray}
\label{ele}
ds^2&=& \left\{1-4G_0\left[ \left(U-T+I^2\right)
\hspace{0.1cm}\ln\left(\frac{r}{r_0}\right)+I^2\hspace{0.1cm}
\ln^2\left(\frac{r}{r_0}\right) \right]\right\} \left(dr^2+ dz^2\right) 
\nonumber \\
&+& r^2\left[ 1-8G_0\left(T-\frac{I^2}{2}\right) - 4 G_0 
\left(U-T-I^2\right) \hspace{0.1cm} \ln \left(\frac{r}{r_0}\right) - 
4 G_0 I^2 \ln ^2 \left(\frac{r}{r_0}\right) \right] d \theta ^2 \nonumber \\
&-& \left\{ 1+ 4G_0 \left[ I^2\hspace{0.1cm} \ln^2 
\left(\frac{r}{r_0}\right) + \left(U-T-I^2\right) 
\hspace{0.1cm} \ln \left(\frac{r}{r_0} \right)\right] \right\} dt^2
\end{eqnarray}
where $U, T$ and $I$ are macroscopic quantities which are defined as 
the energy per unit length,  the tension per unit length and the string's 
current, respectively. 

\section{Formation and Evolution of the Wakes and the Zel'dovich 
Approximation:}

A relevant mechanism to understand the structure formation by cosmic 
strings involves long strings moving with relativistic speed in the 
normal plane, giving rise to velocity perturbations in their wake 
\cite{silk}. Matter through which a long string moves acquires a boost 
in the direction of the surface swept out by the string. Matter moves 
toward this surface by gravitational attraction and, as a consequence,  
a wake is formed behind the string. 
In what follows, we will study the implications of a timelike current 
on the formation and evolution of a wake behind the string which 
generates the metric 
(\ref{ele}). For this purpose, we will mimic this situation with a 
simple model consisting of cold dark matter composed by non-relativistic 
collisionless particles moving past a long string. In order to make a 
quantitative description of accretion onto wakes, we will use the 
Zel'dovich approximation, which consists in considering the Newtonian 
accretion problem in an expanding universe using the method of linear 
perturbations. 

To start with, we first compute the velocity perturbation of massive 
particles moving past the string. If we consider that the string is 
moving with normal velocity $v_s$ through matter, the velocity perturbation 
can be calculated with the help of the gravitational force due to metric 
(\ref{ele}):
\begin{eqnarray}
\label{boost}
u & = &  8\pi G_0 U v_s\gamma \nonumber \\
&  + &   \frac{\pi G_0}{v_s\gamma}\left[2\alpha(\phi_0)^2
\left(U+ T +I^2\right)-\left(U- T -I^2\right)-2\ln
\left(\frac{r}{r_0}\right)I^2\right] 
\end{eqnarray}
with $\gamma = (1-v_s^2)^{-1/2}$. 
The first term in (\ref{boost}) is equivalent to the relative 
velocity of particles flowing past a string in general relativity. The 
other terms come as a consequence of the scalar-tensor coupling of the 
gravitational interaction and the superconducting properties of the string.

Let us suppose now that the wake was formed at $t_i > t_{eq}$. The physical 
trajectory of a dark particle can be written as 
\begin{equation}
\label{traj}
h(\vec{x}, t) = a(t) [ \vec{x} + \psi(\vec{x}, t)] 
\end{equation}
where $\vec{x}$ is the unperturbed comoving position of the particle 
and  $\psi(\vec{x}, t)$ is the comoving displacement developed as a 
consequence of the gravitational attraction induced by the wake on the 
particle. Suppose, for simplification, that the wake is perpendicular to 
the $x$-axis (assuming that $dz=0$ in the metric (\ref{ele}) 
and $ r = \sqrt{x^2 + y^2}$) in such a way that the only non-vanishing 
component of $\psi$ is $\psi_x$. Therefore, the equation of motion for a 
dark particle in the Newtonian limit is 
\begin{equation}
\label{newton}
\ddot{h} =  - \nabla_h \Phi
\end{equation}
where the Newtonian potential $\Phi$ satisfies the Poisson equation
\begin{equation}
\label{poisson}
\nabla_h^2 \Phi = 4\pi G_0 \rho
\end{equation}
where $\rho(t)$ is the dark matter density in a cold dark matter universe. 
For a flat universe in the matter-dominated era, $a(t) \sim t^{2/3}$. 
Therefore, the linearised equation for $\psi_x$ is
\begin{equation}
\label{psi}
\ddot{\psi} + \frac{4}{3t}\dot{\psi} - \frac{2}{3t^2}\psi = 0
\end{equation}
with appropriated initial conditions: $\psi(t_i) = 0$ and $\dot\psi(t_i) 
= -u_i$. Eq. (\ref{psi}) is the Euler equation whose solution is easily found
\[
\psi(x,t) = \frac{3}{5}\left[\frac{u_i t_i^2}{t} - u_i t_i 
\left(\frac{t}{t_i}\right)^{2/3}\right]
\]
Calculating the comoving coordinate $x(t)$ using the fact that 
$\dot{h} = 0$ in the ``turn around"\footnote{The moment when the 
dark particle stops expanding with the Hubble flow and starts to 
collapse onto the wake.}, we get
\begin{equation}
\label{comoving}
x(t) = - \frac{6}{5} \left[ \frac{u_i t_i^2}{t} - u_i t_i 
\left(\frac{t}{t_i}\right)^{2/3}\right]
\end{equation}
With the help of (\ref{comoving}) we can compute both the thickness 
$d(t)$ and the surface density $\sigma(t)$ of the wake \cite{vil}. We 
have, then, respectively (to first order in $G_0$) 
\begin{eqnarray}
\label{thick}
d(t) & \approx & \frac{12}{5} \left(\frac{t}{t_i}\right)^{1/3} 
\left\{ 8\pi G_0 U v_s\gamma  +    \frac{\pi G_0}{v_s\gamma} 
\left[ 2\alpha(\phi_0)^2\left(U+ T +I^2\right)-\left(U- T -I^2\right)-
2\ln\left(\frac{r}{r_0}\right)I^2\right] \right\} \nonumber \\
\sigma(t) & \approx & \frac{2}{5} \frac{1}{v_s\gamma t} 
\left(\frac{t}{t_i}\right)^{1/3} \left[
8U (v_s\gamma)^2 + 2\alpha(\phi_0)^2 \left(U+T+I^2\right)- 
\left( U-T-I^2\right) - 2\ln\left(\frac{r}{r_0}\right)I^2  \right] 
\end{eqnarray}
where we have used the fact that $\rho(t) = \frac{1}{6\pi G_0 t^2}$ for a 
flat universe in the matter-dominated era and that the wake was formed at 
$t_i \sim t_{eq}$. Clearly, from eq. (\ref{thick}) we see that the presence 
of the current makes the accretion mechanism by wakes to  diverge. 

\section{Conclusion:}

Inclusion of a current in the internal structure of a cosmic string could 
drastically change the predictions of such models in the microwave background 
anisotropies. In particular, a current of a timelike-type could bring some 
divergences leading to some unbounded gravitational effects \cite{peter2}.
In this work we studied the effects of a timelike-current string in the 
weak-field 
approximation for the mechanism of 
wakes formation. 
For this purpose, we carried out an investigation of the mechanism 
of formation and evolution of wakes in this framework,  showing the explicit 
contribution of the current to this effect. 

Wakes produced by moving strings can provide an explanation for 
filamentary and 
sheetlike structures observed in the universe. A wake produced by the 
string in one 
Hubble time has the shape of a strip of width $\sim v_s t_i$. With 
the help of the 
surface density (\ref{thick}) we can compute the wake's linear mass density, 
say $\tilde{\mu}$,
\begin{equation}
\sigma(t) \approx  \frac{2}{5} \frac{1}{\gamma t} 
\left(\frac{t}{t_i}\right)^{1/3} \left[
8U (v_s\gamma)^2 + 2\alpha(\phi_0)^2 \left(U+T+I^2\right)- 
\left( U-T-I^2\right) - 2\ln\left(\frac{r}{r_0}\right)I^2  \right] 
\end{equation}
If the string moves slower or if we extrapolate our results to earlier epochs, 
we see that the logarithmic term would bring divergences and the mechanism of 
forming wakes, at least for this model, would break down.

\section*{Acknowledgements}

ALNO would like to thank CAPES for a PhD grant. This work was 
partially supported by 
PROCAD/CAPES.

\end{document}